# Heat meets light on the nanoscale

Svetlana V. Boriskina*, Jonathan K. Tong, Wei-Chun Hsu, Bolin Liao,
Yi Huang, Vazrik Chiloyan, Gang Chen

*Department of Mechanical Engineering, Massachusetts Institute of Technology,
Cambridge, MA 02139, USA; *email: sborisk@mit.edu*

**Abstract**

We discuss the state-of-the-art and remaining challenges in the fundamental understanding and technology development for controlling light-matter interactions in nanophotonic environments in and away from thermal equilibrium. The topics covered range from the basics of the thermodynamics of light emission and absorption, to applications in solar-thermal energy generation, thermophotovoltaics, optical refrigeration, personalized cooling technologies, development of coherent incandescent light sources, and spinoptics.

## 1. Thermodynamics of light and heat

Control and optimization of energy conversion processes involving photon absorption and radiation require detailed understanding of thermodynamic properties of radiation as well as its interaction with matter[1–5]. Historically, thermodynamic treatment of electromagnetic radiation began over a century ago with Planck applying the thermodynamic principles established for a gas of material particles to an analogous 'photon gas'[1]. In particular, he showed that thermodynamic parameters such as the energy, volume, temperature, and pressure can be applied to electromagnetic radiation, reflecting the dual wave-particle nature of photons. In this section, we will review the general thermodynamic principles governing photon propagation and interaction with matter, as well as the evolution of theoretical understanding of these phenomena since original work by Planck[2,6–16].

## 1.1 Photon statistics, temperature and chemical potential

Photons are gauge bosons for electromagnetism and are characterized by their energy $\hbar\omega = hc/\lambda$ and momentum $\hbar\mathbf{k}$, where $h = 2\pi\hbar$ is the Planck's constant, $c$ is the vacuum speed of light, $\lambda$ is the photon wavelength, $\omega$ is the angular frequency, and $\mathbf{k}$ is the photon wavevector[17,18]. Photons can also have angular momentum, including both spin and orbital components[19–21]. The former is associated with the circular polarization of the electromagnetic waves, while the latter originates from electromagnetic phase gradients. Owing to the weakness of inelastic photon-photon scattering process, photons do not exchange energy and thus do not reach thermal equilibrium among themselves during propagation. To account for energy conservation in the interactions of photons with matter, which may lead to the thermalization of a 'photon gas,' the concept of temperature $T$ can be additionally introduced. At thermal equilibrium at temperature $T$, the mean occupation numbers of photons obey the Bose-Einstein statistics and are defined as $\tilde{N}(\omega,T) = \left(\exp(\hbar\omega/k_B T) - 1\right)^{-1}$, where $k_B$ is the Boltzmann constant. The electromagnetic energy density of radiation in a material per unit frequency $\omega$ is defined via the Planck's formula as $U = \hbar\omega \cdot \tilde{N}(\omega,T) \cdot D(\omega)$,[1,2] while the intensity of thermal radiation from the material surface defined based on the power flow, per unit projected area, per unit solid angle, per unit frequency, is $I_\omega = \upsilon_g \cdot U/4\pi$ [22] for isotropic radiation. The parameters $D(\omega)$ and $\upsilon_g$ are the photon density of states (DOS) and group velocity, respectively. In the case of the free-space radiation, the well-known expressions for these parameters are $D = \omega^2/\pi^2 c^3$ and $\upsilon_g = c$. The total power radiated from a blackbody surface per unit area (emissive power), can be found by integrating $I_\omega$ over the whole frequency ($\omega$) and angular ($\varphi, \theta$) ranges: $I = \iint \cos\theta \cdot I_\omega d\omega d\Omega$, where the factor $\cos\theta$ accounts for the surface view factor (and is known as the Lambert's cosine law). For a blackbody source at temperature $T$ emitting into the vacuum, $I = \sigma T^4$, where $\sigma$ is the Stefan–Boltzmann constant.

In general, the number of photons in a photon gas is not conserved. However, light emission and absorption processes involve interactions of photons with other (quasi)particles such as electrons, plasmons, excitons, polaritons, etc., and these interactions obey the conservation laws for energy, momentum and angular momentum. Thus, the number of photons created or annihilated during these interactions cannot always be unrestricted. As a result, photons may derive not only their temperature but also chemical potential[9–11,23–26] from these interactions. However, the classical form of the Planck's law is only applicable to the blackbody radiation generated by incandescent sources and characterized by a zero photon chemical potential $\mu_\gamma = 0$. This formalism cannot properly describe other non-equilibrium absorption-emission processes such as luminescence, lasing, gas discharge, laser cooling or Bose-Einstein condensate[27] formation.

The introduction of the concept of the photon chemical potential enables establishing relations between the number of photons and the number of other quasiparticles photons interact with. These interactions can be modeled in analogy with chemical reactions (i.e., equilibrium is reached when the sum of the chemical potentials, including that of photons, is zero). For example, the distribution of photons emitted at temperature $T$ as a result of recombination of electrons in the conduction band and holes in the valence band of a semiconductor takes the modified form $\tilde{N}_\mu(\omega, T, \mu_\gamma) = \left(\exp\left((\hbar\omega - \mu_\gamma)/k_B T\right) - 1\right)^{-1}$, where $\mu_\gamma = \mu_e + \mu_h$. $\mu_e$ and $\mu_h$ are the chemical potentials of the electron and holes in the emitter.[9,11,28–30] It should be noted that these chemical potential values can only be established after charge carriers reach thermal quasi-equilibrium among themselves within each energy band, forming the quasi-Fermi energy levels for the electron ($F_e$) and hole ($F_h$) distributions in the conduction and valence bands, respectively, so that $\mu_\gamma = F_e - F_h$.

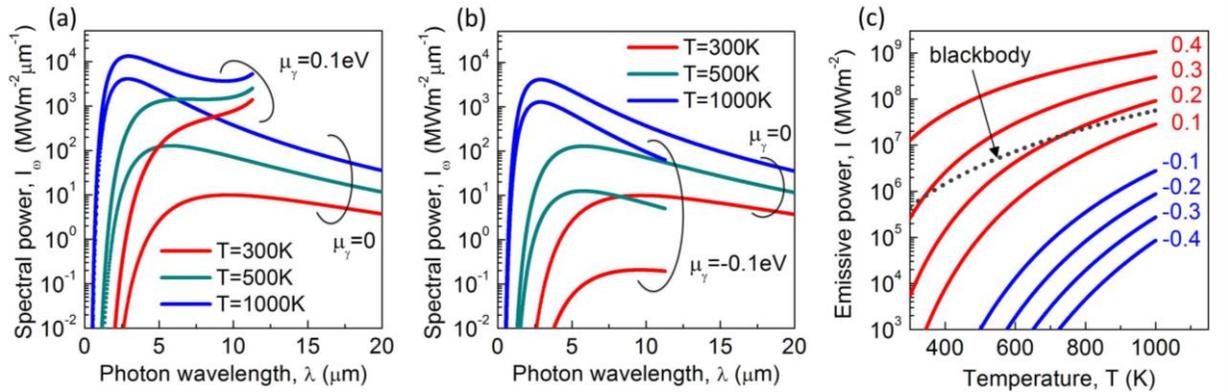

**Figure 1. Chemical potential of light.** (a,b) Heat flux spectra as a function of wavelength and temperature for pure thermal emitters ($\mu_\gamma$=0, do) and luminescent emitters with a photon chemical potential $\mu_\gamma$=0.1eV (a) and $\mu_\gamma$=-0.1eV (b). The emission cut-off energy (defined by the material bandgap) is 0.11eV in both cases. (c) Stefan-Boltzmann law for photons emitted with a chemical potential: emissive power as a function of temperature and chemical potential ($\mu_\gamma$ values are shown as labels). The Stefan-Boltzmann law for a blackbody emitter is shown as a gray dotted line for comparison.

The chemical potential of photons can be introduced by several methods, including photoexcitation, electrical injection, and parametric coupling[31]. For example, photoexcitation of charge carriers in a semiconductor results in the formation of the separate quasi-Fermi levels for electrons and holes, and emission of photons carrying positive chemical potential. This is a typical situation for the operation of photovoltaic (PV) cells[29,32] (see section 1.2) and the efficiency enhancement mechanism in thermohotonic cells and up-converters[33–36] (see section 3.2). Upon substituting the modified photon distribution function into the Planck radiation law, we can observe that photons with a positive chemical potential carry higher energy per photon state than thermally emitted photons at the same emitter temperature. This effect is illustrated in Fig. 1a, which compares photon spectra from sources at varying temperatures and with varying positive chemical potential. Another important distinction from the blackbody spectrum is that the spectra of photons emitted with a chemical potential have a low-energy cut-off, which is defined by e.g., the electronic bandgap of the emitter material (Fig. 1a). If under the external bias applied to the emitter material $F_h$ becomes larger than $F_e$, photons with negative chemical potential are emitted, which carry less energy than the photons in the blackbody spectrum (Fig. 1b). This situation is typical for the operation of thermoradiative[37] cells (see section 1.2), hot-electron energy converters[38] and optical rectennas[39,40].

**1.2 Photon entropy and light energy conversion limits**

Another important thermodynamic characteristic of light is photon entropy[2,3,9,26,41–43]. Understanding the entropy of photons helps to establish the fundamental upper limits for the processes involving conversion of light energy into work and vice versa, including photovoltaics, light generation, and optical refrigeration[3,15,29,44–49]. For a photon flux with a given photon DOS and a distribution function $N(\omega)$, the expression for the entropy per photon state has the following form: $S_\omega = k_B \cdot D(\omega)[(1+N)\ln(1+N) - N \ln N]$, which is applicable to both equilibrium and non-equilibrium cases[2,6]. At the thermal equilibrium, the entropy reaches its maximum and the photon occupation numbers obey Bose-Einstein statistics, $N = \tilde{N}$. The entropy flux leaving the surface of a blackbody source can be calculated via integration over photon energy and direction of light propagation, and equals $4/3 \cdot \sigma T^3$ [2,6]. This expression deviates from the conventional definition of the entropy flow for thermal processes driven by heat conduction, $Q/T$, assuming the heat flow $Q$ is equivalent to the total integrated photon power flow $I = \sigma T^4$. Figure 2a compares the entropy content (i.e., a ratio of the entropy flux to the power flux, $S/I$) of radiation for varying temperatures and chemical potentials with that corresponding to heat conduction ($S/I = T^{-1}$). It can be clearly seen that the entropy content of blackbody emission is higher than that of heat conduction. However, as the chemical potential increases, the entropy of radiation decreases and tends to zero when the lasing condition ($\mu_\gamma \to \hbar\omega$) is reached[9,28]. The low entropic content of lasers and other coherent sources makes possible optical refrigeration, i.e., anti-Stokes fluorescent cooling,[50–52] which is discussed in more detail in section 3.2. Fig. 2a also shows that high-temperature emission is characterized by lower entropy content than the low-temperature one for any value of the chemical potential of photons.

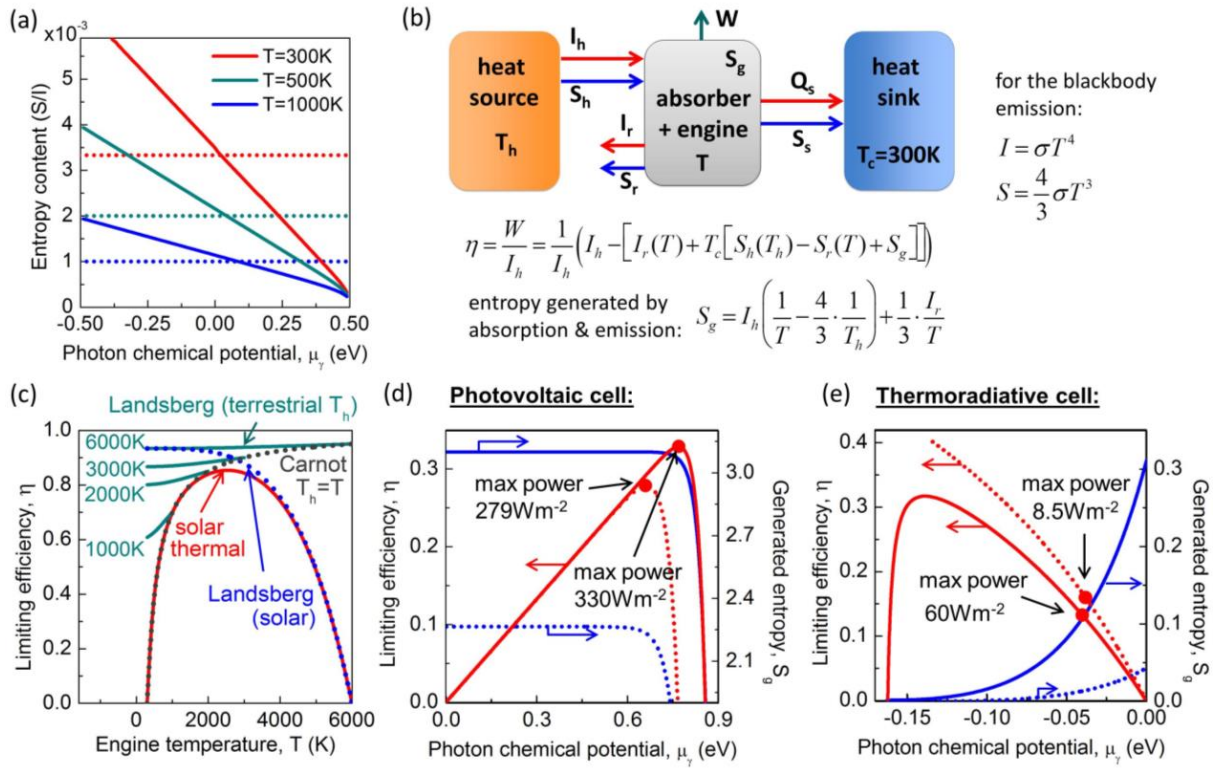

**Figure 2. Entropy of light and optical energy conversion limits.** (a) The entropy content of light, defined as the ratio of the spectral entropy flux to the spectral power flux, as a function of the emitter chemical potential and temperature. The corresponding dotted lines for each temperature denote the entropy content for heat delivered at the same temperature via heat conduction ($S/I = T^{-1}$). (b) Schematic of an energy-conversion device and a general expression for its conversion efficiency. (c) Efficiency limits for ideal engines when heat is delivered from the source to the engine and from the engine to the environment via conduction, i.e. Carnot limit (gray dotted line) or via both radiation and conduction, i.e. Landsberg limit (blue dotted line and teal solid lines). The red solid line shows the limiting efficiency of the solar-thermal energy conversion that accounts for the entropy generation in the processes of photon absorption and emission. (d,e) Limiting efficiencies (red lines) and irreversible entropy generated in the converter (blue lines) of a PV cell with bandgap $E_g$=1.1eV(d) and a TR cell with $E_g$=0.2eV at $T$=500K (e). (d) Solid lines are for the PV cell at $T$=300K, and dotted lines are for the PV cell at T=400K. Red circles are the points of the maximum power generation in the PV and TR cells. (e) Solid lines are for the TR cell emitting photons of all frequencies, and the dotted lines are for a TR cell with a near-monochromatic emission at 0.2eV and a bandwidth of $10^{-5}$eV.

From the above discussion it follows that depending on the origin or radiation, light can be either worse ($S/I > T^{-1}$) or better ($S/I < T^{-1}$) source of energy to extract work from, compared to heat conduction. This has important consequences in the maximum energy conversion efficiency achievable in different types of engines, depending on whether energy is delivered to and from the engine via heat conduction or radiation. A schematic of an engine together with a heat source and a heat sink is shown in Fig. 2b. The heat is delivered to the engine via radiation and the entropy is transferred to the heat sink via heat conduction. The engine emits photons back to the environment at temperature $T$, which is lower than the temperature of the heat source $T_h$. We assume that the heat sink is at temperature $T_c = 300K$, and model the sun as a blackbody emitter with a temperature of $T_h = 6000K$. The general expression for the engine efficiency is also shown in Fig. 2b, which depends on the temperatures of individual elements and on the mechanisms of the energy delivery to and entropy rejection from the engine.

In particular, if heat is delivered to an engine via blackbody radiation ($I_h$) and the engine rejects entropy to the environment via blackbody radiation ($S_r$) and heat conduction ($S_s$), the heat can be converted to work at a maximum efficiency that is less than the classical Carnot efficiency limit[53]. This efficiency is known as the Landsberg efficiency[3,44] (Fig. 2c, blue dotted line), which is calculated under the assumptions that the heat source is the sun and there is no irreversible entropy generation in the engine ($S_g = 0$). Landsberg efficiency reaches its maximum value if $T = T_c$, taking a more familiar form: $\eta_L^S = 1 - 4/3 \cdot T \cdot T_h^{-1} + 1/3 \cdot T^4 \cdot T_h^{-4}$. For the sun as the heat source, $\eta_L = 93.3\%$, representing the upper bound of the solar energy conversion efficiency. For comparison, the efficiency of the ideal Carnot engine[53] reaches 95% efficiency for $T_h = 6000K$ (gray dotted line in Fig. 2c). It has recently been noted that the ultimate efficiency of photon energy conversion can exceed the Landsberg limit in the case of a terrestrial heat source, which can absorb and recycle the heat re-emitted by the engine, yielding $\eta_L^T = W/(I_h - I_r)$[15]. Landsberg efficiency for terrestrial sources at different temperatures $T_h$ is shown in Fig. 2c as teal lines, which reach the Carnot limit when $T = T_h$, and exceed solar Landsberg efficiency for $T_h = 6000K$ and $T > T_c$.

However, the Landsberg limit can only be approached in time non-reciprocal systems[26,54] due to the entropy generation resulting from a local non-equilibrium between photon absorption and emission rates in the absorber material[10,26]. The entropy generated in the processes of the photon absorption and thermal re-emission is $S_g = I_h \cdot (T^{-1} - 4/3 \cdot T_h^{-1}) + 1/3 \cdot I_r \cdot T^{-1}$ [28,55]. Accounting for this entropy generation, the solar energy conversion efficiency in an engine with a blackbody absorber takes a more general form: $\eta = (1 - T^4 \cdot T_h^{-4}) \cdot (1 - T \cdot T_h^{-1})$,[26,56] which is plotted as a function of the engine temperature in Fig. 2c (red solid line). This efficiency reaches its maximum value of 85.3% at $T \approx 2500K$ for a blackbody absorber. However, high efficiencies can be reached at lower temperatures of the absorber if selective surfaces are used (see section 2.1).

The entropy generation in photon energy converters stems from the power mismatch between the incoming radiation and the spontaneous emission of the absorber, and can be mitigated by allowing more degrees of freedom to tailor the spontaneous emission. One example is using the infinite-junction solar cells, where each junction independently interacts with incoming photons at each frequency[28,29]. Another approach is using a stack of two (or more) independently-operated PV cells of the same material but different thicknesses[57]. Although not explicitly discussed in the original publication[57], this strategy enables tuning of two independent photon chemical potentials in individual cells to match the incoming radiation and suppress entropy generation.

To maximize the generated power, engines typically have to be operated away from their point of the lowest entropy generation. For example, in the engines that convert solar energy into an electrical current collected at voltage $V$ (e.g, PV cells), entropy generation is minimized in the open circuit voltage regime, yet to maximize power output, the cell needs to operate at a lower voltage (Fig. 2d). This reduces the chemical potential of re-emitted photons (Fig. 2d, red lines), and increases the irreversible entropy generated in the cell (Fig. 2d, blue lines). The inability of a PV cell to utilize the photon energy in excess of the PV cell voltage (together with the loss of the photons with the energies below the bandgap) results in the limiting efficiency of a single-junction PV cell – known as the Shockley-Queisser limit[29,32] – to deviate significantly from the maximum efficiency discussed in Fig. 2c. It also results in the PV cells heating, which further reduces cell efficiency due to increased energy loss through re-radiation (see Figs. 1a,c), and forces cell operation at an even lower voltage to mitigate this loss (dotted lines in Fig. 2d, $T = 400K$).

Finally, efficiency of some energy converters such as hot-electron cells[38] and thermoradiative (TR) cells[37] is maximized if they are operated at a negative voltage to reduce radiative losses at high temperatures. As illustrated in Fig. 2e, TR cells – to which heat is delivered via conduction and the entropy is rejected via radiation – reach their highest efficiency when operated under large negative voltage, which minimizes radiation and reduces irreversible entropy generation in the cell. However, the highest power is generated under operation at a smaller negative voltage, and is accompanied by the irreversible entropy production. The limiting efficiency of a TR cell can be further increased if it is only capable of emitting near-monochromatic low-energy radiation (dotted lines in Fig. 2e), albeit at the price of the reduced power generation.

**1.3 The role of electron and photon densities of states and the photon angular momentum**

In addition to temperature and photon chemical potential, the intensity and spectrum of radiation emitted by an object can also be modified by manipulating the photon DOS[14,58–66]. Furthermore, most materials do not emit as blackbodies, and an energy-dependent 'gray body' factor typically needs to be introduced to account for intrinsic material properties, which can be tailored by material engineering. For example, for a gray body semiconductor exhibiting an electronic bandgap and thus emitting photons carrying chemical potential, the emission rate is typically calculated via the van Roosbroeck-Shockley equation: $R(\omega) = D(\omega) \cdot \alpha(\omega) \cdot \tilde{N}_\mu(\omega, T, \mu_\gamma)$ [67–69]. This equation is valid for emitters exhibiting a quasi-thermal equilibrium within their respective electronic bands, including excited and ground state electrons, and also for $\mu_\gamma = 0$.

It should be noted that the absorptance coefficient $\alpha(\omega)$ is the rate of photon absorption under the condition of detailed balance between the emission and absorption process, i.e., under a certain quasi-equilibrium electron and hole distributions. As such, unlike the case of the blackbody emission, $\alpha(\omega)$ depends not only on the photon energy and temperature, but also on the rate of the carrier excitation, e.g., photoexcitation or electrical injection. This can be shown by calculating the emission rate by applying Einstein's theory of spontaneous and stimulated emission, $R(\omega) = D(\omega) \cdot \sigma(\omega) \cdot f_u (1 - f_l)$, which is proportional to the product of the absorption rate $\sigma(\omega)$, the probability of the excited electron state being occupied $f_u$ and the ground state being unoccupied $(1 - f_l)$ [68,69]. $f_u$ and $f_u$ are the Fermi-Dirac distributions of charge carries in the emitter material. The absorption rate $\sigma(\omega)$ is calculated under the condition that the upper state is empty and the lower state is full, and thus is uniquely defined by the available charge carrier DOS inside the material and by the transition selection rules imposed by momentum conservation[68,69]. Accordingly, the material absorption rate can be tailored by engineering the electron density of states, e.g., via electron confinement effects. This can yield highly spectrally-coherent emission spectra from quantum dots and wells[16,70–72] as will be discussed in section 2.3. Furthermore, as the emission rate depends on the rate of the carrier excitation, $\alpha(\omega) = \sigma(\omega) \cdot (f_l - f_u)$, the emission spectra can be also manipulated via a combination of increased temperature and either optical pumping or electrical injection[15,73,74] (see section 3.2).

Finally, the photon DOS and group velocity can significantly deviate from their free space values in structured photonic environments, with a profound effect on the spectral characteristics of the emitter. The photon DOS defines the number of states in the momentum space between $k$ and $k + dk$ per unit volume and solid angle $\Omega$ that are available for a photon to occupy in 3D space as: $D(k)dk = k^2 \Omega dk / (2\pi)^3$. DOS as a function of photon energy is expressed via the photon dispersion relation: $D(\omega) = \Omega/(2\pi)^3 \cdot k^2(\omega) \cdot dk/d\omega$ [17,18,63]. It can be seen that high DOS values can be reached for *high-momentum optical states* and/or for *states with flat dispersion characteristics*.

The easiest way to increase the momentum of an optical state, and thus its DOS, is by increasing the refractive index $n$. The dispersion relation for an isotropic bulk material with refractive index $n$ has a well-known form $|\mathbf{k}| = n\omega/c$, which is shown in Fig. 3a for $n = 1$ (dotted line). The photon DOS of an *isotropic* bulk material with index $n > 1$ for both light polarizations is enhanced compared to vacuum, $D_n(\omega) = n^3 \omega^2 / \pi^2 c^3$. The radiative intensity is in turn proportional to $n^2$ since radiative transport requires multiplication with the group velocity of light, $c/n$. This effect can be utilized to maximize light absorption for PV applications[22,66,75,76] and to design high-index radiation extractors for fluorescent[51] and thermal[77,78] emission. In anisotropic media, photon DOS has a more complicated expression, taking different values for ordinary and extra-ordinary rays.[79]

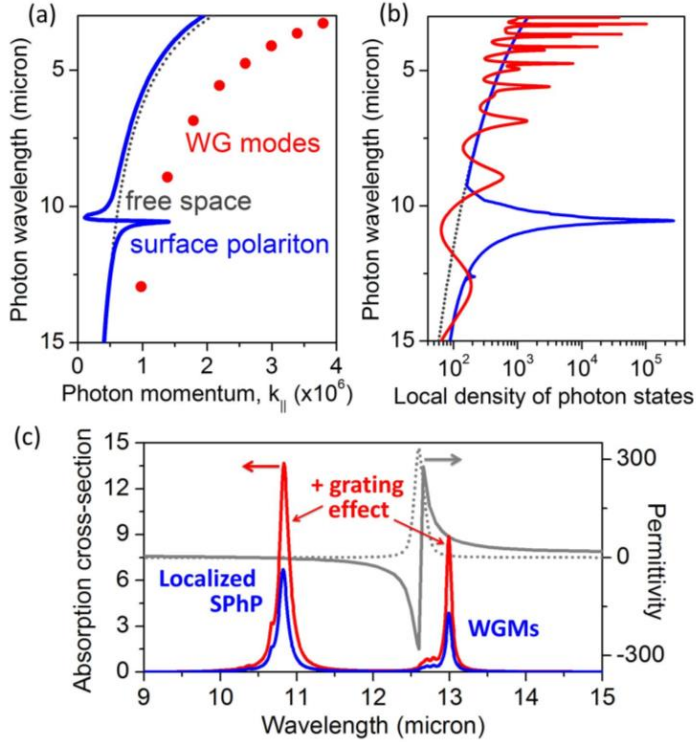

**Figure 3. LDOS engineering with confined optical modes having high photon momentum.** (a) Dispersion characteristics of quantized multiple-degenerate whispering gallery modes in a 5 micron diameter $TiO_2$ microsphere (red dots) and of a surface phonon polariton mode on a SiC-air interface (solid blue line). (b) LDOS frequency spectrum on the surface of the microsphere (red solid line) and on the SiC-air interface (blue solid line). The free space photon dispersion and the free-space photon DOS are shown for comparison as dotted gray lines in (a) and (b), respectively. (c) The absorption efficiency of an 800nm diameter SiC microsphere (blue solid line) and of a periodic chain of 10 microspheres (solid red line) as a function of photon wavelength. The gray lines show the spectral dispersion of the dielectric permittivity for SiC (Re(ε) – solid line, Im(ε) – dotted). The resonant peaks correspond to the excitation of localized surface phonon-polariton modes at ~10-11micron and of trapped whispering gallery modes at ~13micron.

As the photon DOS is inversely proportional to the light group velocity $\upsilon_g = (dk/d\omega)^{-1}$, trapped or guided modes with high photon momenta and/or flat dispersion can have DOS significantly exceeding that of bulk materials[14,58–66,80–82]. As a result, radiative intensity from photonic structures supporting such states can be resonantly enhanced beyond the $n^2$ limit of bulk dielectrics. In particular, resonant effects associated with the excitation of surface polariton waves[61,63,83,84] can be used to strongly modify a material's absorptance and emittance.

Surface polaritons are electromagnetic surface modes existing in metals and polar materials, such as silver, gold, silicon carbide (SiC), and silica ($SiO_2$). Surface phonon polariton (SPhP) modes result from the hybridization of photons and transverse optical phonons, while surface plasmon polaritons (SPPs) are formed owing to the hybridization of photons and volume plasmons – collective oscillations of free electrons. Figure 3a shows the dispersion characteristic of a SPhP mode on a SiC-air interface. The flat region of the dispersion curve corresponds to the excitation of the SPhP at the interface, and is characterized by a strong resonant enhancement of the local density of photon states (LDOS), as shown in Fig. 3b[61,63]. The high longitudinal momentum associated with surface polariton modes is a manifestation of the angular momentum of photons involved in the circulating powerflow along the material interface[85–87]. Although the high-longitudinal-momentum photon states only exist in the sub-wavelength region near the interface, they can be utilized to enhance thermal emission provided that the emitter and the extractor (or the absorber) are coupled through the near field[12,63,72,83,88–96]. Near-field coupling between SPP or SPhP modes on multiple material interfaces also enables engineering photonic metamaterials characterized by hyperbolic photon dispersion, known as hyperbolic metamaterials (HMMs).[63,65,97–100] HMMs are highly anisotropic, and the spectral energy density of photons in HMMs drastically differs from that predicted by Planck's blackbody theory. The Stefan-Boltzmann law for hyperbolic media has been shown to be a quadratic (rather than fourth order) function of the temperature[79]. HMMs are also characterized by high photon DOS in broad frequency bands, which can be utilized for engineering solar absorbers[101,102] and thermal emitters both in the near and far field[63,103,104] (see section 5.1).

Confined quantized photon states in optical nanostructures can be also used to enhance and tailor energy extraction via photon emission in both near and far field regimes. These photon states include guided modes in thin films[62,75,80,105–107], fibers and nanowires[108–111], trapped modes and localized polariton modes of optical micro- and nano-cavities[59,62,81,112–115], and photonic crystal defect modes.[116,117] Thermal emission extraction via engineering photon LDOS in micro/nanostructures builds upon the obvious parallels with the electronic DOS engineering in quantum wells, wires, and dots[62,105].

Figure 3a shows the dispersion characteristics of high-momentum whispering gallery modes (WGMs) trapped in a 5-micron $TiO_2$ microsphere (i.e. a thermal dot) by total internal reflection[112,118]. WGMs are characterized by a high angular photon momentum $k_{ang} = (l(l+1))^{1/2}$, defined by the mode quantum number $l$ [119]. The linear tangential momentum $k_\parallel = (l(l+1)/r^2)^{1/2}$ reaches a high value on the sphere surface. Its role in the generation of the high photon LDOS is analogous to the role of the high in-plane photon momentum of surface polaritons on planar interfaces. The photon LDOS on the microsphere surface is shown in Fig. 3b as the red line and features sharp peaks corresponding to the WGM excitations at discrete frequencies. Each WGM has a frequency degeneracy of $2l+1$, due to several allowed azimuthal momentum quantum numbers $m=-l,...l$ for each angular momentum number $l$ [112,118]. This degeneracy – which may be removed in micro- and nano-cavities of more complex geometries or due to near-field electromagnetic coupling[114,120,121] – further increases the LDOS of the high-$k$ WGM states.

Although WGM states exhibit high LDOS, curvature of a sphere or cylinder results in total internal reflection that is not as ideal as for planar interfaces. As a result of this curvature, it is possible for these modes to leak energy into the far field, thus enabling radiation extraction and external coupling. In fact, this very argument can also be applied in the case of particles supporting localized surface polariton modes as well. In Fig. 3c, we plot the absorption efficiency for a SiC microsphere, which we calculated using Mie theory[59,84] as the ratio of the sphere's scattering cross-sections to its geometrical cross-section (blue solid line). The two sharp peaks observed in Fig. 3c correspond to the excitation of the localized SPhP modes at the frequencies within the Restrahlen band of SiC characterized by a negative dielectric permittivity, and trapped WGMs outside of this band. From the detailed balance principle, the absorptance spectra in Fig. 3c can be used to calculate the corresponding emittance from the sphere into the angular element $d\Omega_\mathbf{k}$ as $\varepsilon_\mathbf{k}(\omega)=\alpha_\mathbf{k}(\omega)$ [59]. The absorptance and emittance can be further enhanced via far-field electromagnetic coupling between microspheres arranged into a square-lattice array with the period on the order of radiation wavelength (red solid line).

Figure 3c illustrates that individual high-DOS wavelength-scale absorbers can exhibit absorption cross-sections that are larger than the corresponding geometrical cross-sections[84,122], which demonstrates the possibility of emittance that is higher than a blackbody[59] as well as the possibility for generating spectrally coherent thermal light. Indeed, thermal emission from wavelength-scale particles and nanowires[84,110,123] is among the few known examples of far-field radiation exceeding the blackbody limit. The emittance or absorptance of extended arrays of individual nanoscale emitters, which is calculated per unit surface area rather than the localized emitters' surface area, cannot exceed the blackbody limit[60]. Nevertheless, long-range electromagnetic coupling in such arrays can enable highly localized heating via photon absorption, and, by reciprocity, enhanced light extraction from nanoscale thermal emitters.

It is important to note that DOS modification via trapped photon states (e.g., WG modes) does not necessarily rely on the excitation of surface modes in the emitter material, and thus offers more flexibility in their spectral design. As seen in Fig. 3, trapped photons states can be excited at frequencies much higher than those typical for SPhP excitation, which offers applications in optical energy conversion schemes, such as thermophotovoltaics[62,105,124].

## 2. Heat is the new light: manipulation and harvesting of thermal emission

Spontaneous emission, including thermal emission, typically yields incoherent and unpolarized light. The ability to control the frequency, spectral coherence and directionality of thermal and thermally-enhanced fluorescent radiation would be a significant benefit for applications including solar energy harvesting, radiative cooling, sensing, and spectroscopy. In this section, we review the state-of-the art and remaining challenges associated with spectrally shaping the spontaneous emission via engineering of both the electron and photon densities of states. DOS engineering has been successfully used in the past to either enhance or inhibit light absorption, fluorescence rates, Raman scattering efficiency, and near-field energy transfer for applications such as on-chip communications and sensing[115,118,125–132]. However, the broadband nature of thermal emission and the temperature dependence of the photon state occupancies make photon DOS engineering for heat extraction more challenging.

### 2.1 Selective surfaces for solar-thermal energy generation

The incorporation of spectral selectivity in radiative absorbers and emitters can benefit many applications, most notably, in the field of solar-thermal energy conversion. Solar-thermal energy converters with blackbody receivers suffer from re-emission losses at high temperatures[133]. In Fig. 4a we illustrate this challenge by comparing the spectral photon energy fluxes for AM1.5[134] solar radiation and thermally emitted infrared radiation from a blackbody at high temperatures. Typically, solar-thermal receivers require a high level of solar concentration to overcome these losses. However, achieving high optical concentration, e.g. by using a large heliostat field, may not only be costly, but also introduces additional energy losses, which reduce overall system efficiency and can have negative environmental impact.

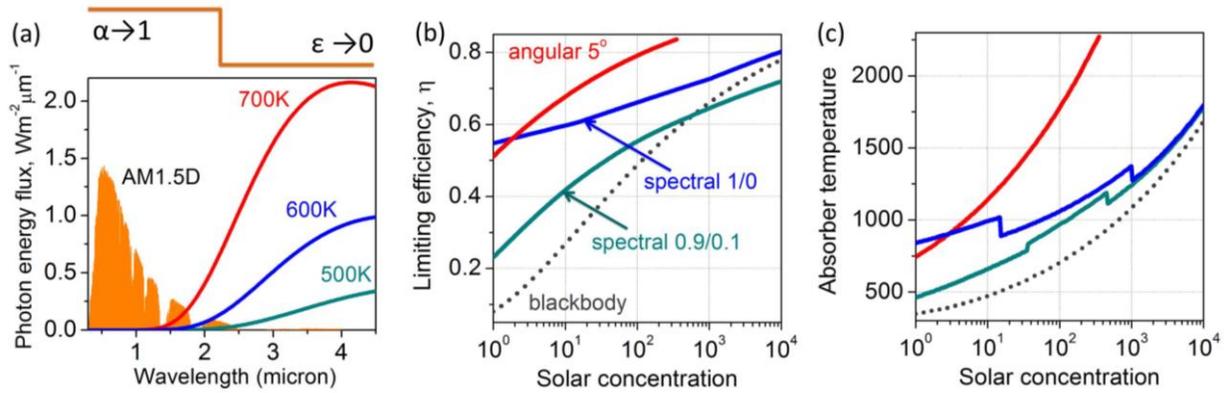

**Figure 4. The role of spectral and angular selectivity in solar-thermal energy conversion.** (a) Frequency spectra of terrestrial solar radiation (AM1.5D, orange) and of thermal radiation of blackbody emitters at varying temperatures. The top inset shows the ideal spectral absorptance/emittance characteristics of the solar receiver surface. (b) The limiting efficiency of solar-thermal energy converters with a blackbody (gray dotted line) and selective (solid lines) absorbers as a function of solar concentration. The engines are assumed to convert thermal energy at the Carnot efficiency. For spectrally-selective absorbers, the transition frequency between high and low emittance levels (labeled as ratios α/ε) is optimized for each concentration ratio. Angularly-selective absorber has no spectral selectivity (1/1), yet is only able to absorb and emit photons in the narrow angular range of +/-5 degrees from normal. (c) The corresponding equilibrium temperature of the absorbers with the optimized transition point between high and low emittance levels discussed in (b) as a function of solar concentration.

To suppress re-emission losses, a receiver with a step-wise emittance, as shown in the inset to Fig. 4a, needs to be engineered[15,135–138]. Ideally, such a receiver would feature a complete absorptance across most of the solar spectrum, and zero emittance at lower frequencies overlapping with the thermal emission spectra. As illustrated in Fig. 4b, even a non-ideal receiver with 90% solar absorptance and 10% infrared emittance can increase the maximum possible efficiency of the solar energy conversion. The better functionality is achieved due to the ability to reach higher receiver temperatures at low solar concentrations (Fig. 4c). The values in Figs. 4b,c for each solar concentration have been calculated by optimizing the transitional frequency separating high-absorptance and low-emittance spectral regions to yield the highest equilibrium temperature of the receiver. Due to the presence of atmospheric absorption bands in the solar spectrum, the equilibrium temperature does not vary smoothly with the solar concentration, as evidenced by the kinks in Fig. 4c[139].

At this point, it should be mentioned that the receiver designs presented above with high solar absorptance and low thermal emittance do not violate the Kirchhoff's law. Although the common corollary of the Kirchhoff's law is typically assumed to be that a good absorber has to be a good emitter, and by contrast a poor absorber, a poor emitter, it is important to realize that terrestrial absorbers operate at temperatures much lower than the temperature of the Sun. Thus, these *receivers are not in thermal equilibrium with solar radiation*. They however obey the general form of the Kirchhoff's law, also known as the principle of detailed balance, $\varepsilon(\omega,\theta,\varphi,T) = \alpha(\omega,\theta,\varphi,T)$, where $\varepsilon(\omega,\theta,\varphi,T)$ is the directional spectral emittance, $\alpha(\omega,\theta,\varphi,T)$ is the directional spectral absorptance, and the angles $\theta,\varphi$ specify light propagation direction.

Absorbers with angular selectivity can offer even higher energy conversion efficiencies for solar-thermal energy conversion applications (red solid line in Fig. 3b). Unlike conventional (Lambertian) emitters with isotropic emission patterns, angular-selective emitters can only absorb/emit light within a narrow angular range[140]. As sunlight illuminating the Earth surface is highly collimated, such receivers can still achieve perfect absorptance of solar energy[141]. To calculate the efficiency shown in Fig. 4b, the angular-selective receiver is assumed to exhibit no spectral selectivity. The combination of both spectral and angular selectivity can yield an even higher efficiency at low solar concentrations. Angular selectivity would also benefit a myriad of other applications including solar receivers, thermophotovoltaic (TPV) energy converters, and energy-efficient directional incandescent light sources.

Spectral and angular selectivity can be achieved by using external filters or by designing the absorber to emit selectively in different directions. Several possible approaches to achieve thermal emission with angular selectivity are illustrated in Fig. 5. The first approach shown in

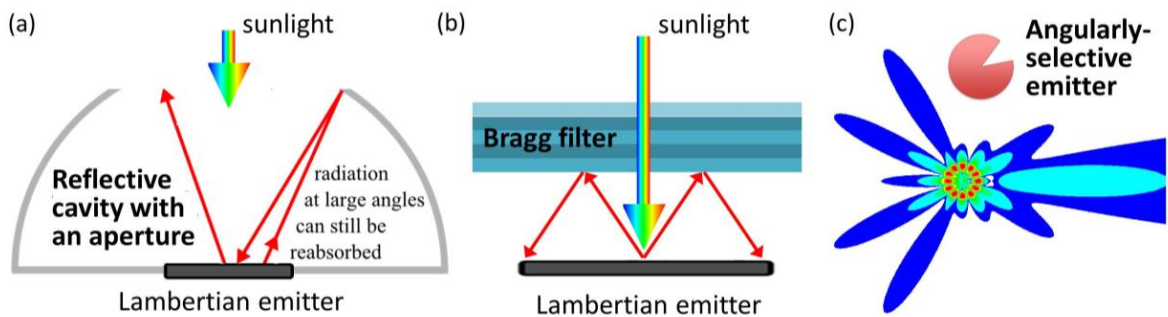

**Figure 5**. **Possible approaches to realize angular selectivity of a thermal emitter**. (a) The emitter is embedded into a reflective cavity with a narrow-angle aperture for thermal radiation[142–144]. (b) The emitter is covered with an angular-selective filter, which only allows photon transmission within a narrow angular range. (c) Asymmetrical nanoscale emitters can have high angular selectivity without the use of external reflective optical elements[148,150].

Fig. 5a is based on a reflective cavity with an aperture[142–144]. Thermally-emitted photons propagating in the direction close to the normal to the emitter surface escape the cavity through the aperture. However, photons emitted at larger angles are reflected back to the emitter and re-absorbed, thus dramatically reducing thermal emission losses. The aperture also allows sunlight to illuminate and heat the receiver. Similarly, multi-layered dielectric Bragg filters with spectrally- and angle-selective properties recently demonstrated by Soljacic and colleagues (Fig. 5b) can also be used to recycle photons emitted at angles that differ from the chosen direction of emission[140] or at frequencies outside of a select frequency band[145]. This approach also allows combining spectral and angular selectivity by designing the filter to have omnidirectional reflectance within chosen frequency range(s). The selective filter can also be attached directly to the emitter surface[137,146,147].

Asymmetric nano- and micro-scale objects supporting localized optical resonances can also exhibit highly non-isotropic angular emission patterns as illustrated in Fig. 5c for the case of a notched WGM microdisk. Local perturbations in symmetrical emitters can help to match the momentum of the high-$k$ trapped modes to the low-$k$ propagating modes in free space, thus facilitating far-field extraction of the optical energy. In particular, Boriskina has previously predicted that localized indentations can be used to generate highly directional light emission from WGM microcavities[148,149], which has been subsequently proven experimentally[150].

Finally, long-range electromagnetic coupling effects in gratings or photonic crystals can provide momentum matching of propagating photons in free space to high-$k$ polariton modes supported by metals, doped semiconductors or polar dielectrics. This long-range coupling yields highly directional thermal emission from gratings at select frequencies corresponding to the excitation of high-LDOS surface modes.[151,152] Some examples of directional thermal sources based on gratings will be discussed in section 2.3. Gratings can also be designed to provide spectral selectivity for light absorption and thermal emission. This functionality is illustrated in Fig. 6a, which shows calculated absorptance spectra for single- and multiple-periodic hafnium (Hf) gratings. Hf was chosen for this demonstration for use in a high temperature solar receiver as this metal forms a stable native oxide layer, which is advantageous for high-temperature operation in air rather than in vacuum, and can also be used to further improve solar absorption. It can be seen in Fig. 6a that the metal surface spectral selectivity can be improved by patterning the metal surface, and is tunable by the pattern geometry and symmetry properties. The emittance of the patterned surface drops sharply in the long-wavelength spectral range above 3 micron, which can help prevent re-emission losses at high temperatures. Other patterned metal surfaces – including tungsten (W), tantalum (Ta) and gold (Au) – have been demonstrated to provide similar or even better spectral selectivity[138,153,154].

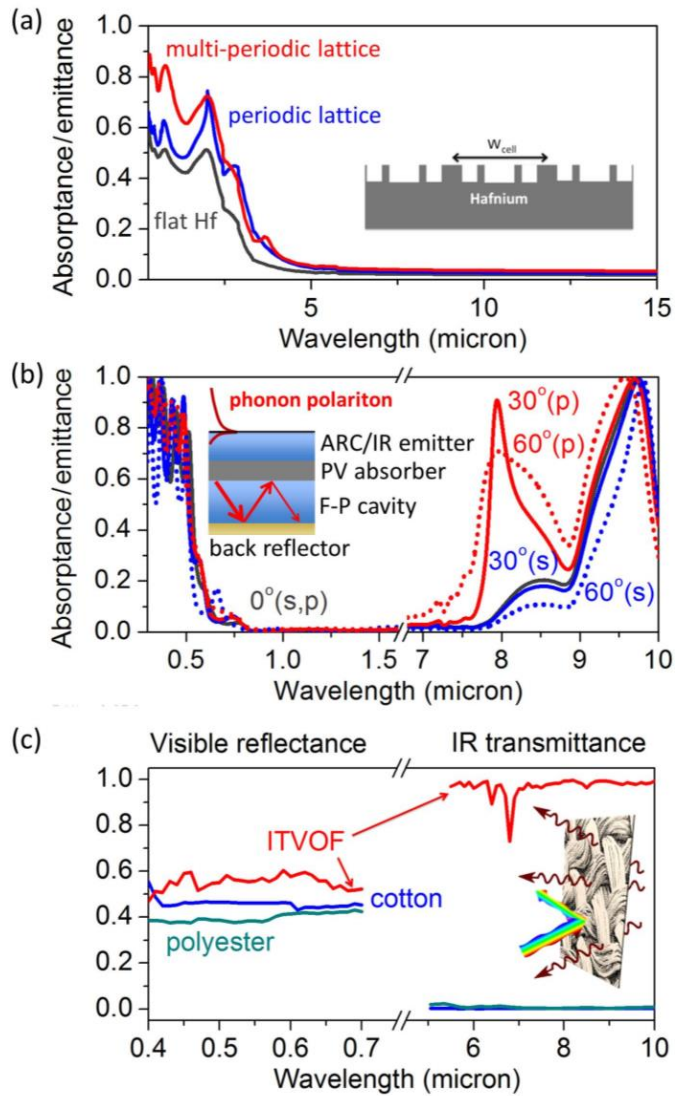

**Figure 6. Nanophotonic solutions for spectral selectivity of thermal emittance.** (a) Absorptance spectra of a metal (Hf) receiver with a flat surface (gray line), a periodically-modulated surface (blue line), and a surface with a multi-periodic pattern shown in the inset (red line). (b) Absorptance/emittance spectra of a multi-layered solar receiver with a 50-nm amorphous Si as a PV absorber, a silver backreflector, an absorber-mirror spacer that acts as a Fabry-Perot cavity, and an overlayer made of silica. The spectra are calculated at different angles to normal and for both s- and p-polarizations. The top silica layer serves as both an anti-reflection coating for visible photons and a thermal emitter for IR photons. (c) IR-transparent visible-opaque fabric (ITVOF)[169]: visible reflectance and IR transmittance spectra for conventional fabrics (cotton, blue, and polyester, teal) as well as the optimally-designed ITVOF (red). The inset is a schematic illustrating the ITVOF performance.

Despite the good performance of such structures for creating thermal emission selectivity, the fabrication of large-scale micro- and nano-scale patterns at the emitter surface can be a technologically challenging and costly endeavor. To address this challenge, the development of spectrally-selective emitters with simpler geometries is highly desirable. One recently proposed approach to achieve spectral selectivity and suppress thermal emission is to cover a solar receiver with a layer of transparent silica aerogel. The aerogel provides efficient thermal insulation[155] and helps to reduce the IR emittance of the receiver, offering a simple planar technological solution to improve the efficiency of a solar-thermal energy converter. Another possible solution to achieving spectral selectivity, which will be discussed in the next section, is to place thin film emitters on top of optimally-designed substrates or embed them into planar Fabry-Perot optical cavities. Selective absorbers and emitters designed using this approach have been successfully demonstrated both theoretically and experimentally[107,156–159].

**2.2 Mesoscale resonant structures for passive radiative cooling**

Unlike solar-thermal receivers, photovoltaic (PV) solar converters need to operate at low temperatures. However, low-energy photons cannot be converted to electricity in PV cells; instead, they contribute to parasitic heating of the device. To alleviate this problem and to increase efficiency of PV cells, it was recently proposed by Shanhui Fan and colleagues to combine high solar absorption with high thermal emission in the infrared spectral range[160,161]. Radiative cooling of PV cells and other devices requires development of nanostructures and materials with drastically different electromagnetic response in the visible and infrared frequency ranges. For this application, high absorptance in the visible range needs to be accompanied by the high emittance in the infrared region. The IR emittance should ideally peak in the frequency range corresponding to the atmospheric transparency window in 8 to 13 micron wavelength range[162–166]. A selective emitter radiating strongly in this range is thus exposed to the clear sky, enabling the use of the outer space as a very low temperature heat sink. Radiative cooling can be achieved by choosing an optimal combination of material and optical properties.

This approach to tailor the spectral properties of solar receivers is illustrated in Fig. 6b, which shows the simulated absorptance and emittance spectra of a multi-layer absorber with an 80nm-thick layer of amorphous silicon (a-Si) as the active layer for potential PV applications. The 150nm silica ($SiO_2$) spacer underneath the a-Si layer is supported by a gold reflector and thus acts as a planar Fabry-Perot cavity. Constructive interference of incident and reflected light within this cavity can lead to an increase of solar absorption. The absorptance in the visible frequency range drops sharply at the frequency corresponding to the a-Si material bandgap. In turn, radiative cooling of this structure can be achieved as a result of IR emittance enhanced by the SPhP mode excitation within the Restrahlen band of a 700nm silica layer on top of the a-Si film, which spectrally overlaps with the atmospheric transparency window. This layer simultaneously serves as the transparent anti-reflective coating in the visible range, which helps to enhance solar absorption in a-Si. In the structure reported in Fig. 6b, we have numerically optimized the thicknesses of each individual layer to increase IR emittance and to simultaneously enhance solar absorption. More complicated photonic crystal structures have been designed to achieve even higher IR emission for the PV cell cooling applications[166,167].

Personalized cooling technology is another example of an application where optical spectral selectivity can be beneficial. Conventional personal cooling is typically achieved through heat conduction and convection. However, past studies have shown the human body to be a very efficient emitter of IR radiation[168], which suggests radiative cooling can also be an effective cooling mechanism. Most conventional fabrics are opaque to IR radiation (see Fig. 6c) and block thermal emission from the body to the environment. Ideal fabrics for personalized cooling applications thus need to exhibit spectral selectivity in their transmittance characteristics. They should enable IR transmission to directly pass through clothing, thus maximizing radiative cooling, while ensuring that the fabric is opaque at visible wavelengths. We have recently demonstrated that resonant electromagnetic effects in microfibers can provide a basis for the development of a new wearable technology for personalized cooling[169].

The fabrics with the required spectrally selective transmittance spectra were designed using a combination of optimal material composition and structural photonic engineering. Synthetic polymers that support few vibrational modes were identified as candidate materials to reduce intrinsic material absorption in the IR wavelength range. Individual fibers were designed to be comparable in size to visible wavelengths to achieve strong light scattering and thus remain optically opaque to the human eye. At the same time, they are too thin to support trapped photon states at IR wavelengths. This reduces the photon LDOS in microfibers and minimizes the interaction of the fabrics with IR light, resulting in high transmission in this spectral range.

As illustrated in Fig. 6c, the designed fabrics are indeed opaque for the visible light yet transparent for the infrared thermal radiation emitted by the human body. Compared to conventional personal cooling technologies, these fabrics can provide a fully passive means to cool the human body regardless of the person's physical activity level. The anticipated effect of the new optical-thermal polymer fabrics (tailored into a short sleeve shirt and pants) will be the increase in the body cooling rate by at least 23 W[169]. This would enable thermally regulated interior environments, such as offices or homes, to raise the summer HVAC temperature set points from 75°F to 79°F, resulting in significant energy savings[170].

## 2.3 Nanophotonic incandescent sources: coherent, polarized and twisted thermal light

Spectral and angular selectivity of radiation is a signature of the temporal and spatial coherence of the thermal emitter, respectively. Many of the above approaches to increase coherence of the thermally-emitted light have been pursued both theoretically and experimentally, as illustrated in Fig. 7. Coherent thermal emitters can potentially be used as tunable light sources for sensing and spectroscopy, especially in the far-infrared range, where hardly any coherent sources are available[171]. They can also play an important role in optical communications and solar and waste heat energy harvesting, e.g., via a thermophotovoltaic energy conversion process[49,138,172–174] or optical rectification[39,40]. The degree of coherence for a radiation source with a known spectral distribution $I(\omega)$ can be calculated as $\gamma(\tau) = \int_0^\infty I(\omega) \cdot \exp(-i\omega\tau)d\omega$, where $\tau$ is the time delay between interfering light sources, which is determined by the optical path difference $\Delta$ as $\tau = 2\Delta/v_g$ [175–177]. The coherence length is typically defined as the optical path difference at which the degree of coherence reaches 10%. The coherence length of thermal blackbody radiation scales inversely with temperature of the source, and equals about 0.6 micron for the sunlight[178] and about 4 micron for the radiation from an incandescent emitter at $T\sim775K$[177].

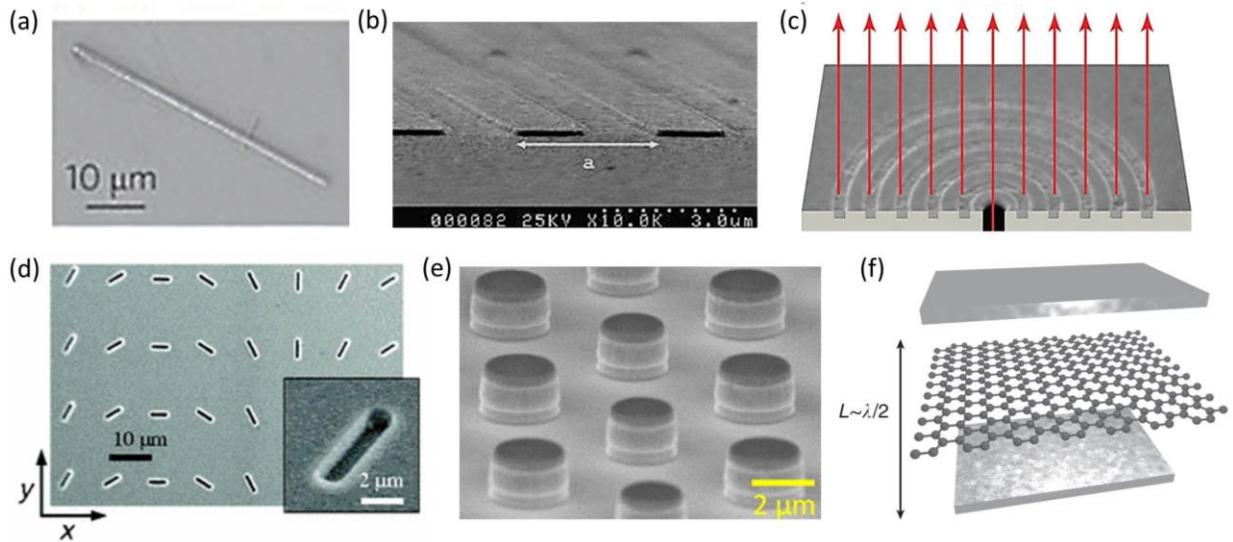

**Figure 7. Nanophotonic realizations of thermal emitters with partial spectral and spatial coherence based on photon and electron confinement effects.** (a) A SiC single nanowire emitter (reproduced with permission from[110], ©NPG), (b) a periodic grating of tungsten nanoribbons (reproduced with permission from[151], ©OSA), (c) a plasmonic bulls-eye grating (reproduced with permission from[183], ©NPG), (d) an array of nanorod antennas with spatially-varying local anisotropy axis (reproduced with permission from[194], ©APS), (e) a periodic photonic crystal lattice with embedded GaAs quantum wells (reproduced with permission from[71], ©AIP), and (f) a graphene sheet embedded into a planar Fabry-Perot cavity (reproduced with permission from[200], ©NPG).

However, resonant enhancement of photon and electron DOS in the emitter via a combination of optical and quantum confinement effects can yield nano- and micro-scale thermal sources with coherence and polarization characteristics of emitted photons deviating significantly from blackbody radiation. This is important for thermal emission manipulation via optical interference, as only [partially] coherent light exhibits interference phenomena, while completely incoherent light sources contribute additively to the intensity at any point of space. Furthermore, the temporal and spatial coherence of the thermal emitter are manifested in the spectral and angular selectivity of emission, which stem from the position-momentum and time-energy uncertainty principles, and can be utilized in the photon energy conversion schemes.

For example, it has been shown by Greffet and others that planar sources supporting surface polariton modes yield quasi-monochromatic near-field radiation with the coherence length larger than the emitted photon wavelength[179]. Thermal sources with geometries that provide matching of the high momentum of these surface modes to the free space radiation may exhibit even longer coherence length than planar near-field emitters made of the same material[180,181]. These coherent far-field directional thermal sources can be shaped as nanowire antennas shown in Fig. 7a[110] or planar gratings of with either parallel[151] or concentric[182,183] grooves as shown in Figs. 7b,c. For example, the coherence length of thermal emission from individual Ti/Pt nanowires of 125 nm in width was experimentally measured to be at least 20μm at $T$=773K, which is much larger than expected for a blackbody emitter[177]. Such thermal sources with sizes on the order or smaller than the wavelength of emitted photons are especially interesting, as they can exhibit *global coherence* if the coherence length is comparable to or exceeds the wavelength. In turn, the coherence length of the near-infrared tungsten grating emitter shown in Fig. 7b is measured to be about 106 micron at $T$=623K[151]. This value exceeds the coherence length of the blackbody emission and is approaching a coherence length of a $CO_2$ laser. Other coherent thermal emitters based on periodic one-dimensional gratings and two-dimensional photonic crystals with emission peaks observed in near-to-far-IR frequency range have been demonstrated experimentally by using metals and polar dielectrics[180,181,184–189].

Thermal emission from coherent nanoscale sources quite often exhibits a significant degree of polarization[190]. For example, radiation from heated thin nanowires supporting localized plasmon and polariton waves is strongly polarized in the direction orthogonal to the wire[191,192]. As the polarization direction of emitted light follows the orientation of nanoscale emitters, thermal emission from nanopatterned surfaces can be tailored by introducing local anisotropy of individual emitters' orientation. As shown in Fig. 7d[181,193,194], local anisotropy can be introduced by gradually rotating the direction of nanorod antennas patterned on a planar substrate. It has been shown experimentally by Hasman and colleagues that thermal radiation emitted by such a surface exhibits photon momentum characteristics associated with the emission from a revolving medium. This phenomenon is known as the angular Doppler effect,[181,193,194] and is a manifestation of the spin-orbit interaction of emitted light. Spin-orbit interactions are also a basis for other interesting optical effects such as photon analogs of spin-Hall,[195–197] Magnus[198], and Coriolis[199] effects, which paves the road to many exciting applications of coherent nanoscale thermal sources in the emerging field of spinoptics.

Finally, thermal emission can be controlled by simultaneous engineering of photon LDOS and electron DOS in the material. The latter modifies intrinsic absorption characteristics of the material and can be achieved by using quantum confinement effects, for example, in quantum wells as shown in Fig. 7e[16,70,71]. In this structure, quantum-well intersubband transitions strongly modify material absorption properties, which – in combination with photon DOS modification due to a periodic arrangement of nanoscale emitters – results in a dramatic narrowing of the spectral and angular range of emitted radiation. Likewise, monolayer materials such as graphene demonstrate unusual electronic and optical properties due to the electron confinement in the monolayer plane. Embedding a graphene layer inside a planar optical microcavity as shown in Fig. 7f provides a way to modify graphene emission characteristics and to achieve highly spectrally-selective thermal emission[200].

The devices shown in Figs. 7e,f can be pumped by electrical current injection, which heats up the emitter due to scattering of injected electrons. If the optical modal structure of the thermal source inhibits radiation of long-wavelength photons, the thermal energy accumulates inside the emitter in the form of the energy of phonons and 'hot' charge carriers, which leads to the emitter self-heating. As a result, emitter can reach significantly higher temperatures under the same input power and thermal management conditions as compared not only to the blackbody reference but also to the same material in the absence of photon confinement effects[16,70,200]. Furthermore, at high temperatures, the thermal conductivity of the suspended graphene is significantly reduced due to the strong Umklapp phonon–phonon scattering[201], which also suppresses lateral heat dissipation and leads to spatial localization of hot electrons. As a result, the efficiency and brightness of thermal emission is further increased, making possible realization of thermal graphene sources emitting in the visible range[202].

## 3   Thermal up- and down-conversion of photon energy

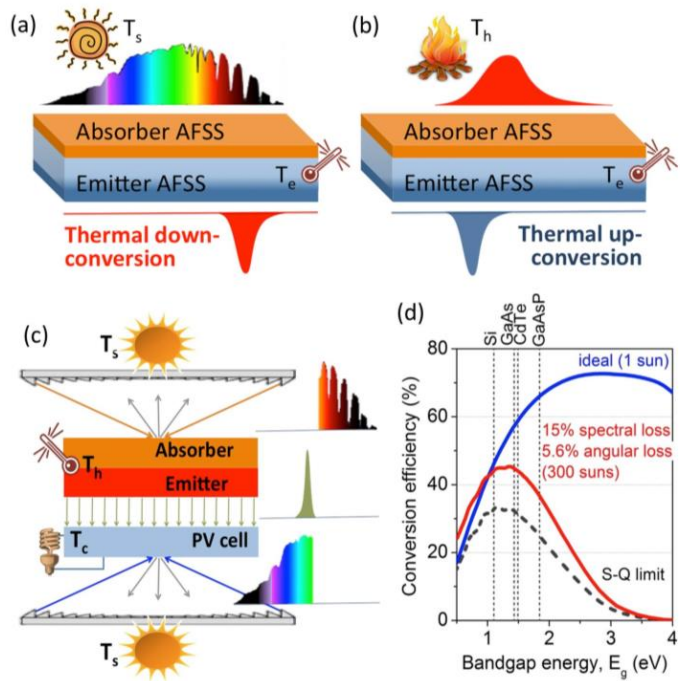

**Figure 8. Thermal up- and down-conversion of photon energy.** (a,b) Schematics of thermal down- (a) and up-converters (b) based on the use of angular- and spectrally-selective surfaces for the control of thermal emission. (c) Conceptual scheme of a solar energy conversion platform based on a PV cell enhanced by the thermal up-converter[208]. (d) Theoretically-predicted maximum achievable efficiency of the device shown in (c) as a function of the PV cell electron bandgap.

Thermal energy storage and recycling inside emitters with spectral- and angular-selective characteristics can be used to achieve thermal up- and down-conversion of photon energy via the process of photon absorption and subsequent controlled thermal re-emission. This is illustrated in Figs. 8a,b, which schematically show photon absorbers and thermal emitters with angular- and frequency-selective surfaces (AFSS). The absorbed photons are coupled to another surface via heat conduction and re-emitted in desirable frequency range. By adjusting the spectral selectivity of the absorbing and emitting surfaces, either up- or down-conversion process can theoretically be achieved. Solar thermophotovoltaic energy conversion platforms effectively make use of the photon down-conversion process shown in Fig. 8a to convert the broadband solar spectrum to a narrow-band thermal emission spectrum, which peaks at lower photon energy tuned to fit the electron bandgap of a PV cell[138,146,203–207]. Thermal up-conversion of photon energy (Fig. 8b) is more challenging, yet could be highly promising for applications in waste heat harvesting and in development of tunable photon sources.

## 3.1 Thermal up-conversion

We have recently predicted[208] that solar-to-electricity conversion efficiency higher than the conventional Shockley-Queisser limit for PV cells[32] can be achieved in a hybrid platform that combines a single-junction solar cell and a thermal up-converter, as illustrated in Fig. 8c. Within such a hybrid scheme, photons with energies below the bandgap of the PV cell are absorbed by the up-converter, which heats up and re-emits photons with higher energies towards the cell. To make this process possible, both front and back surfaces of the up-converter must have carefully designed angular- and frequency-selective emittance characteristics. The front surface should provide a perfect absorption of sunlight across the solar spectrum, and prevent re-emission of photons at lower frequencies. Additionally, it should reduce re-emission of photons with energies overlapping with the solar spectrum, which requires the surface to exhibit angular as well as spectral selectivity. This can be achieved, for example, via the use of an external cavity with an aperture or a selective filter as shown in Figs. 5a,b. In turn, the emitter surface should only allow emission of photons with energies above the PV cell bandgap.

The maximum efficiency of a hybrid device utilizing the thermal up-conversion scheme can reach 73% for the ideal selective emittance characteristics (perfect solar absorption and perfect thermal emission blocking) of the up-converter surfaces under illumination by the non-concentrated sunlight (Fig. 8d)[208]. Reaching high up-conversion efficiency requires raising the up-converter temperature, which for the commonly used photovoltaic materials – e.g., Si, GaAs, CdTe, and GaAsP – nevertheless was predicted to lie within practically achievable 900-1600K range. Non-ideal up-converter surfaces that allow for reasonable absorption and emission losses still yield limiting efficiency values exceeding 45% for moderate optical concentration of 300 suns (Fig. 8d). The observed increase in the predicted conversion efficiency is a result of better matching of the PV cell bandgap energy to the energy of incoming photons, which reduces irreversible entropy generation in the PV cell. For the ideal angular and spectral characteristics of the up-converter surfaces, the entropy creation rate in the up-converter is much lower than that in the PV cell, which explains high maximum efficiency levels shown in Fig. 8d. However, the increase of the radiation losses from the up-converter with non-ideal characteristics results in the increase of the entropy creation rate in the up-converter (see section 1.2), which is reflected in the drop in the hybrid device efficiency. Successful realization of such hybrid platform could offer new opportunities for reaching higher levels of solar energy conversion efficiency under low levels of optical concentration[15,208].

## 3.2 Laser cooling and heat-assisted luminescence

Anti-Stokes luminescence is another example of a process based on thermally-induced up-conversion of the photon energy. The thermal up-conversion process discussed in the previous section relies on the energy storage as the phonon or free electron energy, accompanied by controlled photon re-emission in thermal equilibrium with the material. The photon up-conversion via anti-Stokes luminescence instead makes use of a non-equilibrium photon emission as a result of radiative recombination of electron-hole pairs assisted by electron-phonon scattering, as schematically shown in Fig. 9a. This process results in the emission of photons carrying chemical potential (see section 1.1) and with energies that are higher than the energy of photons in the optical pump, which forms a basis for optical refrigeration (also called laser refrigeration or anti-Stokes fluorescent cooling)[50–52]. Optical refrigeration can be achieved by irradiating a luminescent material at a frequency on the low energy tail of its absorption band, which is followed by spontaneous emission of more energetic photons in the process of the anti-Stokes luminescence (Fig. 9b). The extra energy of emitted photons is extracted from the lattice phonons, which results in the cooling of the material. The anti-Stokes luminescence can also be observed under material pumping via electron injection[209–212].

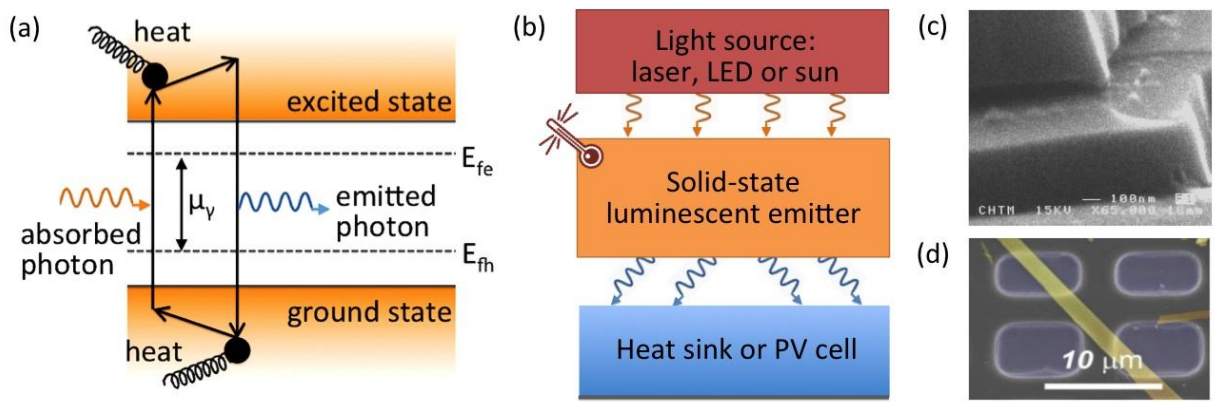

**Figure 9. Heat-assisted luminescence as a basis for optical refrigeration and photon up-conversion.** (a) Energy diagram illustrating the process of the anti-Stokes luminescence under optical pumping. (b) Schematic of the energy exchange between an optical pump, a luminescent material exhibiting anti-Stokes luminescence, and a heat sink, which absorbs up-converted photons. (c,d) Demonstrated photonic solutions for enhancing the efficiency of laser cooling, including a high-index absorber optically-coupled to the luminescent emitter through (c) a nanogap (reproduced with permission from[221], ©SPIE) (c), and (d) a nanobelt cryocooler (reproduced with permission from[222], ©ACS).

Although optical refrigeration was originally proposed by Pringsheim almost a century ago,[50] the progress in this field was stymied for decades, first, by the concerns for possible violation of the thermodynamics second law[213], and later by the lack of high-purity materials. Landau helped to resolve the controversy with the second law violation by proper accounting for the entropy of light emitted in the process of spontaneous isotropic anti-Stokes fluorescence[6] (see section 1.2). In turn, progress in materials engineering yielded materials with high internal quantum efficiency, which enabled successful experimental demonstrations of the optical refrigeration, mostly of rare-Earth-doped grasses[42,51,52], starting with the experiments of Epstein and co-authors[42]. The efficiency of the laser cooling process scales with the material absorption efficiency $\eta_{abs}$ and the external quantum efficiency $\eta_{ext}$ as $\eta_c = \eta_{abs} \cdot \eta_{ext} \cdot (\lambda_p / \lambda_f) - 1$, where $\lambda_p$ is the wavelength of the optical pump, and $\lambda_f$ is the wavelength of the spontaneous anti-Stokes fluorescent emission. Accordingly, a lot of effort in increasing this efficiency has been directed to material engineering in order to increase the pump photon absorption rate and to suppress non-radiative decay due to multi-phonon emission processes. In particular, electron DOS engineering with quantum wells, dots and impurity bands has been explored to increase the cooling efficiency[64,210,214–217]. However, the external quantum efficiency can be strongly affected by the photon DOS in the emitter and its near-field neighborhood, which paves the way to the potential efficiency increase via photonic engineering[51,64,218].

The role that the photon DOS engineering can play in boosting the efficiency of the optical refrigeration is two-fold. First, a fluorescence peak can be blue-shifted by suppressing the photon DOS in a spectral range above the pump wavelength $\lambda_p$, thus decreasing $\lambda_f$. Additionally, an increase of the radiative rate at the blue-shifted emission wavelength could increase the external quantum efficiency $\eta_{ext}$. Photonic crystals[219] and planar surfaces supporting surface-plasmon polariton modes[64,220] have been theoretically proposed by Khurgin as promising means to improve the cooling efficiency, however, they have not been experimentally demonstrated to date. Two successful examples of the nanophotonic structures that improve the efficiency of the optical refrigeration are shown in Figs. 9c,d. The structure in Fig. 9c makes use of the $n^2$ emission rate enhancement (see section 1.3) in a high-index absorber. The absorber is coupled to a luminescent emitter across a nanoscale gap, which provides thermal insulation and enhances photon extraction via frustrated total internal reflection process, i.e., photon tunneling across the gap[51,221]. High-index photon extractors shaped as domes, which are directly attached to the luminescent emitter have also been experimentally demonstrated[51]. In turn, the use of optically-thin nanobelt cryocoolers (Fig. 9d)[218,222] enables suppression of the photon DOS *inside* the luminescent material and thus increases the external quantum efficiency. Instead of getting trapped inside the material and optically recycled until they are re-absorbed, up-converted photons can escape the nanobelt structure, resulting in its temperature reduction. Likewise, photon DOS engineering has potential to improve cooling of solids via the process of spontaneous anti-Stokes Raman scattering[223].

Finally, as the heat-enhanced anti-Stokes spontaneous emission and scattering processes enable photon energy up-conversion, they can be used to increase the efficiency of solar energy conversion processes. One possible realization of a hybrid solar energy converter using heat-assisted photoluminescence has recently been proposed[15,73,74,224–226], and is schematically illustrated in Fig. 9b. In such a converter, solar radiation is absorbed in a low-bandgap photo-luminescent emitter, which is kept at high temperature. The heat sink is replaced with a PV cell with the bandgap energy higher than the bandgap of the emitter. The up-converted photons emitted in the process of the anti-Stokes luminescence gain enough energy to be harvested by the high-bandgap PV cell. This enables converting thermal energy into electrical work at high voltage, which yields enhanced device efficiency. Such a hybrid conversion platform has been theoretically predicted to reach maximum efficiency of 70% at operating temperatures below 1000°C. Operation at even lower temperatures (<500K) can be achieved if the thermal gradient is maintained by a hot electron-hole gas rather than by a hot lattice, such as e.g., in a hot-carrier solar cell operated at open circuit[34]. Recent experiments demonstrated 30% up-conversion efficiency by using a InGaAs/GaAsP quantum well up-converter, where the thermal gradient was maintained by steady-state Auger heating of charge carriers[35].

## 4 Bridging heat conduction and radiation on the nanoscale

As discussed above, tremendous progress has been made in understanding and controlling thermal emission, which resulted in successful demonstrations of coherent thermal sources. However, unless thermal emitters are operated at high temperatures, they are only suitable for low power applications, especially when compared to conventional coherent emission sources such as lasers and light emitting diodes as well as alternative heat transfer mechanisms such as heat conduction and convection. At realistic emitter temperatures, the power output of thermal sources and optical energy converters is low, even if they can theoretically operate at very high efficiencies. One approach to boost the power output relies on storing energy in the form of non-equilibrium excitations, with subsequent emission of photons that carry chemical potential. This enables generation of a higher energy flux than equilibrium blackbody radiation at the same temperature of the emitter (see Fig. 1a). This approach forms the basis for the operation of the hybrid energy converter based on the heat-enhanced anti-Stokes luminescence (Fig. 9b) [15,73,74,224]. Another approach makes use of the dramatic enhancement of the local photon DOS in the near-field of a thermal emitter to increase the radiative power of a thermal source[61]. Past studies have successfully demonstrated both theoretically and experimentally that by separating an emitter and an absorber by distances comparable to the wavelength of radiation, it is possible to enhance radiative transfer by several orders of magnitude beyond the blackbody limit[12,90,91,93–96,99,227–232].

## 4.1 Near-field DOS enhancement & resonant energy transfer

The latter approach is demonstrated in Fig. 10a, which compares the photon energy flux spectrum of a blackbody emitter with the corresponding spectra for a thin-film 'thermal well' emitter[62,105], collected in the far- and the near-field, respectively. It can be seen in Fig. 10a that the photon DOS modification in thin-film emitters can be used to efficiently suppress long-wavelength far-field thermal radiation (blue solid line). The photon DOS suppression below cut-off frequencies of the guided modes trapped in the thin film by the total internal reflection is analogous to the case of the nanobelt cryocooler[218,222] discussed in section 3.2. This spectral selectivity translates into high theoretically-predicted efficiency of TPV platforms using thin film emitters and thin film PV cells[105]. However, the total energy flux delivered by photons from the emitter to the PV cell drops dramatically below the blackbody emitter level, which would reduce the power output of a TPV platform. The observed drop in the power flux is due to an inability of the high-momentum guided modes in the thin film to couple to the free-space propagating modes that carry energy away from the emitter into the far field (top inset in Fig. 10a). In contrast, if the same emitter and the PV cell are coupled through the near field, the radiative energy flux is significantly enhanced and can exceed the blackbody level by orders of magnitude while retaining spectral selectivity (red solid line). The basis for the observed enhancement is the efficient coupling of the high-*k* trapped waveguide modes across the vacuum gap, which results in the enhanced tunneling of emitted photons into the absorber (middle inset in Fig. 10a).

Even more dramatic enhancement of the near-field radiative heat transfer rate can be achieved by using near-field emitters and absorbers supporting surface polariton modes, which exhibit significant photon DOS enhancement at resonance, as explained in section 1.3. Figure 10b presents calculated near-field heat flux spectra between planar emitters and absorbers made of either aluminum zinc oxide (AZO, blue solid line) or SiC (teal solid line) and coupled across a 20nm-wide vacuum gap[72]. Both spectra exhibit resonant peaks due to the electromagnetic coupling of surface plasmon-polariton modes in AZO and phonon-polariton modes in SiC across the vacuum gap (bottom inset in Fig. 10a). These peaks exceed the blackbody limit by orders of magnitude. Figure 10b also illustrates that further radiative heat flux enhancement can be achieved if the emitter and the absorber are shaped as ultra-thin films (compare the red solid line to the blue one). The additional enhancement has been attributed to the significant spectral broadening of the emission spectra due to coupling between SPP modes on both sides of each thin film[72,233]. Similarly, mode-coupling-induced spectrum broadening and enhancement of the radiative heat flux has been predicted and demonstrated for thin-film SPhP emitters[92,93,234–237] as well as for metamaterial[98,103,104,147,186,192,238,239] emitters composed of multiple thin films (or nanowires) supporting polariton modes.

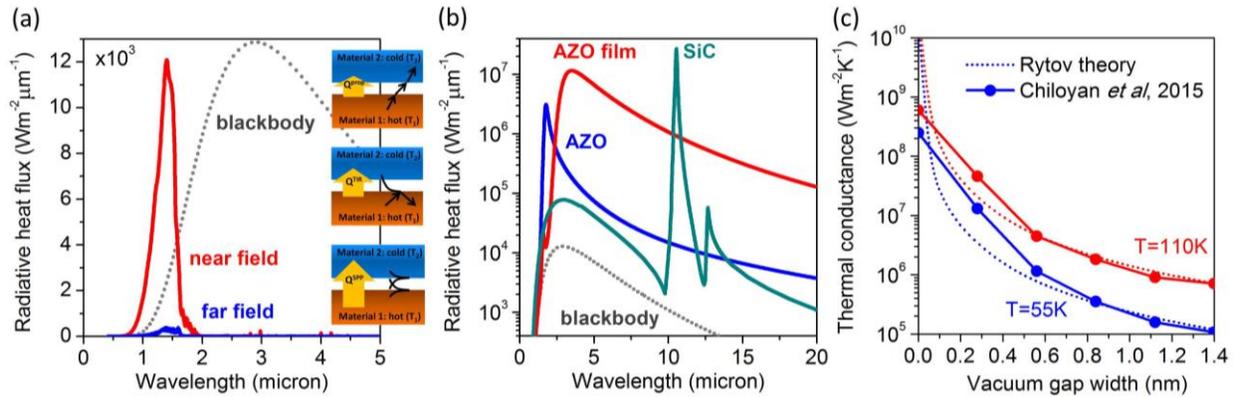

**Figure 10. Near-field effects in the enhancement of the radiative heat transfer.** (a) Spectral radiative heat flux between a thin-film 'thermal well' emitter and a thin-film absorber[105] in both near-field (10nm vacuum gap, solid red line) and far-field (100micron vacuum gap, solid blue line) regimes with respect to a blackbody heat flux at the same temperature (grey dotted line). The emitter is a Ge thin-film of 860nm thickness and an absorber is a GaSb film of 136nm thickness, both of which are supported by perfect back-reflectors[105]. The hot and cold side temperatures are 1000 K and 300 K, respectively. The insets schematically illustrates the three channels used for the radiative heat transfer, from top to bottom: via propagating waves, via photon tunneling due to frustrated total internal reflection, and via near-field coupling of high-LDOS surface polariton modes. (b) The heat flux spectra as a function of wavelength between various emitters and absorbers supporting surface polariton modes. The gap is 20 nm wide, and the hot and cold side temperatures are 1000 K and 360 K, respectively. The blackbody heat flux is shown as the dotted line. (c) Thermal conductance as a function of gap width calculated by using the classical (Rytov) theory[7,8] and a recently developed atomic approach[242].

**4.2 Bridging conduction and radiation**

To date, most of the calculations of both the far-field thermal emission and the near-field radiative energy transfer have relied upon macroscopic Maxwell equations – either exact or under various approximations. Although the continuum electromagnetic models smear out the atomic details, they are applicable in most practical cases. However, these details cannot be ignored when the separation between thermal emitters and absorbers approaches atomic length scales. As a result, the electrodynamic theory developed by Rytov[7,8] to model the near-field radiative heat transfer, which is based on the fluctuation-dissipation theorem[240], diverges as the gap between an emitter and absorber approaches zero (see Fig. 10c), which is unphysical. Accordingly, a mechanistic atomic-scale description of the transfer mechanisms is required for more accurate prediction of energy transfer at very small separations between two bodies.

To develop such an approach, we utilized the Green's function formalism using lattice dynamics with the microscopic Maxwell equations[241]. By using the new approach, we investigated the mechanisms underlying energy transport between two ionic slabs in the near field radiation regime down to contact within a single unified formalism[242]. Figure 10c shows the calculated values of conductance plotted as a function of slab separation for small gaps to show the discrepancy between our approach (dots) and Rytov theory (solid line) near contact. The Rytov theory demonstrates the unphysical divergence near contact whereas our theory is able to yield a finite contact conductance. The new model resolves the zero-gap divergence issue and provides a computational tool to probe the conductance from the near field radiation regime down to contact in a unified manner. Interestingly, the new results show that near-field radiative heat flux at very small gaps can be enhanced even stronger than previously predicted via continuous electromagnetic models.

## 5  Discussion: remaining challenges and outlook

Despite the tremendous progress in understanding various mechanisms of thermal emission manipulation and applying this knowledge to development new device designs, many challenges still remain, both, fundamental and technological.

### 5.1 Progress and challenges in materials engineering and fabrication

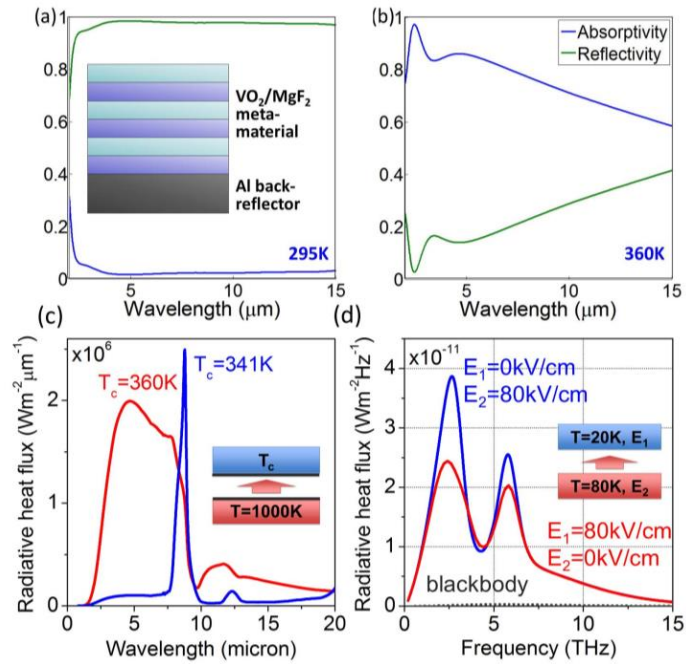

**Figure 11. Dynamic tunability of thermal emission.** (a,b) Absorptance/emittance spectra of a metamaterial structure composed of $VO_2$, $MgF_2$, and an Al back-reflector. At low temperatures, the structure acts as a mirror (a), while at high temperatures it becomes an efficient broadband emitter (b). (c) The near-field heat flux spectra as a function of wavelength between 2 nm-thin $VO_2$ films on $SiO_2$ substrates at an emitter temperature of 1000K and an absorber temperature either at 360K (red line) or 341K (blue line)[72]. The emitter and absorber are separated by a 20 nm gap. (d) Simulated radiative heat flux between ferroelectric materials for a 100nm gap between hot and cold surfaces[260]. Magnitude of spectral heat flux is modulated via external electric field.

For example, fabrication of nanoscale gaps between an emitter and absorber remains a significant technological challenge. As a result, experimental demonstration of high-efficiency near-field TPV platforms is lacking despite many promising conceptual proposals. However, radiative heat transfer between a nanoscale tip and a planar surface can be achieved at very narrow separations. This offers many interesting technological applications, including thermal radiation scanning tunneling microscopy[243–246], high-resolution thermal lithography[89] and heat-assisted magnetic recording (HAMR),[247–249] where nanoscale thermal emitters are used to heat a magnetic recording medium. The progress in this field will also benefit optical refrigeration and photon up-conversion applications, where the figure of merit can be improved by using the near-field fluorescence extraction mechanism[220,250].

The continued progress in materials engineering is also essential for the development of new material platforms with optical properties tuned to yield high photon DOS at various frequencies from the visible to the far-infrared[171,251–253]. In particular, dynamically-tunable thermal emitters are of high interest, with applications to sensing, spacecraft thermal management, and thermal camouflage. Materials that actively alter their thermal emittance through phase change have been investigated for this purpose, most notably, vanadium dioxide ($VO_2$), which undergoes an insulator-to-metal transition at 68°C[254,255]. This leads to the thermal emission reduction if the temperature is increased above the transition temperature, contrary to the blackbody emission trend, and offers applications in infrared camouflage[256]. However, unique properties of $VO_2$ can also be used to design metamaterial emitters that emit strongly at temperatures above the phase transition as illustrated in Figs. 11a,b. The emitter shown in Fig. 11a has a periodic structure composed of thin layers of $VO_2$ and magnesium fluoride ($MgF_2$), which is a transparent dielectric in mid-IR, and is supported by an aluminum (Al) back-reflector. At low temperatures, the multi-layer structure is largely transparent in the infrared, and thus the overall emittance is governed by the Al mirror, resulting in a low emittance (Fig. 11a). In contrast, at high temperatures, $VO_2$ becomes metallic, and the multi-layer structure undergoes a transition to become a hyperbolic metamaterial, which exhibits enhanced emission across the entire IR wavelength range due to its hyperbolic photon dispersion[63,65,97–100] (Fig. 11b). Such metamaterial emitters with variable heat rejection can be useful for terrestrial and spacecraft thermal management applications[257].

$VO_2$ emitters can also be used to achieve strong dynamic modification of the near-field heat transfer, especially in combination with polar materials such as $SiO_2$ or SiC (Fig. 11c). A thin film of $VO_2$ in its metallic phase supports SPP modes, which enhance the near-field radiative heat flux at high temperatures and screen the SPhP mode of the silica substrate (red solid line). When the $VO_2$ film becomes transparent in its insulator phase, it enables the SPhP mode to transfer energy across the nanoscale gap between the emitter and the absorber, resulting in a dramatic spectral transformation of the near-field heat flux (blue solid line). It should be noted that in the case of low-temperature operation, the radiative near-field heat flux from a $VO_2$ surface is dominated by the strong contribution from the SPhP polariton modes supported by the insulator phase of $VO_2$ at low frequencies (above 20 μm in wavelength)[96].

Active tuning of thermal emission via external electric or magnetic fields could enable higher switching rates, but to date, has only been demonstrated in a handful of materials. These include fast modulation of the intersubband emission of the quantum well p–i–n diodes[258] and tuning of the graphene thermal emission[259] via application of gate bias, which varies the carrier density in the emitter material. Emission properties of perovskite-structure ferroelectric materials can also be modulated by application of an external dc electric field, as illustrated in Fig. 11d, and temperature[260]. However, as this process is most effective near the Curie temperature – i.e. phase transition temperature between ferroelectric and paraelectric phases – development of ferroelectric materials with high Curie temperatures is necessary for practical applications. Other types of phase-change or voltage-tunable materials that can be explored for developing tunable thermal emitters include W-VO2 alloys[261], ITO[262], and Ge2Sb2Te5 (GST)[263] just to name a few. Achieving material stability at high temperatures and outside of vacuum environment is also a big challenge that needs to be overcome[15,154,264].

**5.2 Advances and challenges in fundamental understanding and nanophotonic design**

Some fundamental aspects of the light absorption, emission as well as light-to-work energy conversion are still under investigation. In particular, the search is still on-going for the optimum conditions to achieve higher-than-blackbody emittance of coherent nanoscale sources, or, by reciprocity, enhanced resonant absorption at select frequencies. Some success in increasing the absorption cross-sections has been achieved by tuning the geometry and material composition of nanoscale emitters to achieve near-degeneracy of several trapped photon modes of different polarization and angular momentum in a narrow spectral range[265,266]. Metamaterial hyperlenses have been theoretically predicted to yield selective absorber heating and, by reciprocity, enhanced light extraction from nanoscale thermal sources into the far field[78,267,268]. It has also been proposed that extraordinarily large absorption cross sections of nanoscale resonators can be achieved by embedding them into a material with near-zero refractive index[269]. Recent progress in designing and fabricating low-loss zero-index metamaterials operating at various frequency ranges[270–273] makes feasible practical realization of this approach. The role of collective effects[274] caused by the interference between multiple coherent thermal sources in thermal emission enhancement, radiation wave front shaping, and focusing still needs to be revealed[275,276]. In particular, fundamental and technological solutions for thermal emission focusing can help to overcome challenges in realizing the near-field coupling schemes.

Fundamental studies on the entropy and information content of correlated light fields generated by coherent thermal sources are also important[277–279], especially in the processes of light absorption and emission away from thermal equilibrium, when light-matter interactions occur over time scales too short for the thermalization process to take place. This is illustrated by a recent striking demonstration of efficient optical refrigeration via the heat transport between a thermal bath and a non-equilibrium exciton-polariton fluid[280]. Other theoretical studies predict that a modification of a photonic system coupling to the thermal bath via parametric oscillations may create an effective chemical potential for photons even in thermodynamic equilibrium[31]. This offers potential applications in quantum electrodynamics and optomechanics.

Finally, addressing the issues associated with heat removal from the laser gain media require detailed understanding of the energy and entropy of the photons, electrons and phonons in laser systems[41,281–284]. Development of a complete theoretical framework connecting the fundamental atomic physics, thermodynamics, and material science associated with photon absorption, emission and non-radiative thermal transport phenomena at nano/micro and macro scales is necessary to address these issues.

## Acknowledgements

The authors would like to thank Laureen Meroueh, Kenneth McEnaney, Lee Weinstein, Marin Soljacic, Shanhui Fan, Carmel Rotschild, Peter Bermel, Philippe Ben-Abdallah, David McClosky, Calinda M. Yew, Baratunde Cola, and Richard M. Osgood for helpful discussions. This work was supported by DOE-BES Award No. DE-FG02-02ER45977 (for thermal emission tailoring and near-field radiative heat transport), the ARPA-E award DE-AR0000471 (for hybrid full spectrum harvesting), and the 'Solid State Solar-Thermal Energy Conversion Center (S3TEC)', funded by the US Department of Energy, Office of Science, and Office of Basic Energy, Award No. DE-SC0001299/DE-FG02-09ER46577 (for thermophotovoltaic applications).